\documentclass[a4paper]{article}

\PassOptionsToPackage{numbers, compress}{natbib}

\usepackage[preprint]{nips_2018}

\usepackage[utf8]{inputenc} 
\usepackage[T1]{fontenc}    
\usepackage{hyperref}       
\usepackage{url}            
\usepackage{booktabs}       
\usepackage{amsfonts}       
\usepackage{nicefrac}       
\usepackage{microtype}      
\usepackage{ amssymb }
\usepackage{pgfplots}
\usepackage{float}
\usepackage{subcaption}
\usepackage{setspace}
\usepackage{graphicx}

\title{Human Vocal Sentiment Analysis}

\author{
 Andrew Huang , Martin (Puwei) Bao \\
 New York University Shanghai\\
 \texttt{\{\href{mailto:amh877@nyu.edu}{a.huang}, pb1713\}@nyu.edu} \\
}

\begin{document}
\doublespacing
\maketitle
\begin{abstract}
In this paper, we use several techniques with conventional vocal feature extraction (MFCC, STFT),  along with deep-learning approaches such as CNN, and also  context-level analysis, by providing the textual data, and combining different approaches for improved emotion-level classification. We explore models that have not been tested to gauge the difference in performance and accuracy. We apply hyperparameter sweeps and data augmentation to improve performance. Finally, we see if a real-time approach is feasible, and can be readily integrated into existing systems.
\end{abstract}
\section{Introduction}
\subsection{Context}
 Sentiment analysis has become widely prevalent in many areas such as text and images and is useful for deriving meaningful information from otherwise context-less situations. With the rise of personal assistants such as Google Assistant and Amazon Alexa, it becomes important to give meaningful responses to queries based on the users mood. Another area where human speech classification is useful is in product reviews \cite{perez-rosas_utterance-level_2013} . A large percentage of online speech reviews that would previously need to be manually labeled can now be analyzed, which is invaluable to product manufacturers and sales. Vocal sentiment has traditionally been done through feature extraction (pitch, MFCC, formants, etc.) and auditory filters. Recently, more context-free approaches have been taken with machine learning models.
 \subsection{Objective}
\begin{figure}[H]
\caption{}
\vspace{-5cm}
    \begin{subfigure}[b]{\textwidth}\vspace{5cm}
    \centering
    \includegraphics[width=\textwidth]{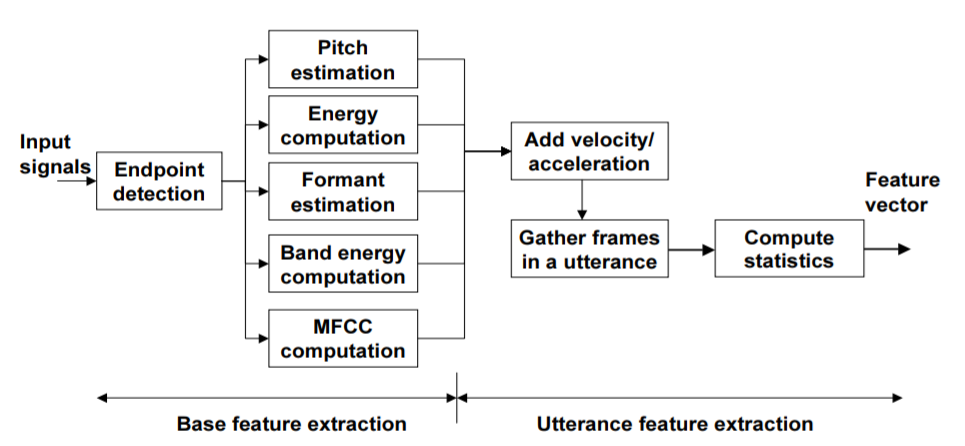}
    \caption{Vocal Feature Extraction}
    \label{figure:fe}
\end{subfigure}
\begin{subfigure}[b]{0.5\textwidth}
    \centering
    \includegraphics[width=\textwidth]{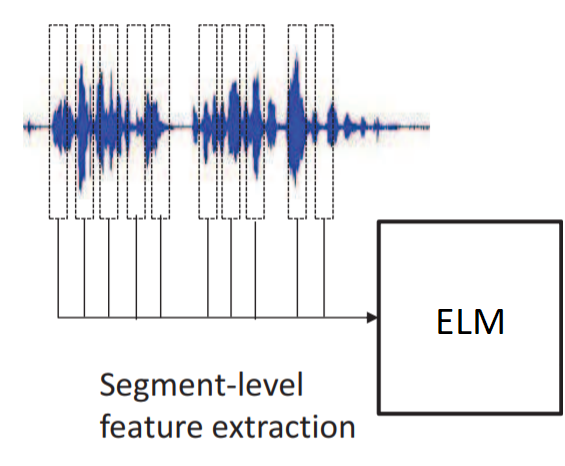}	        \captionsetup{justification=centering}
	\caption{Classification with Segment-level Feature}
	 \label{figure:sfe}
\end{subfigure}
\begin{subfigure}[b]{0.5\textwidth}
    \centering
    \includegraphics[width=\textwidth]{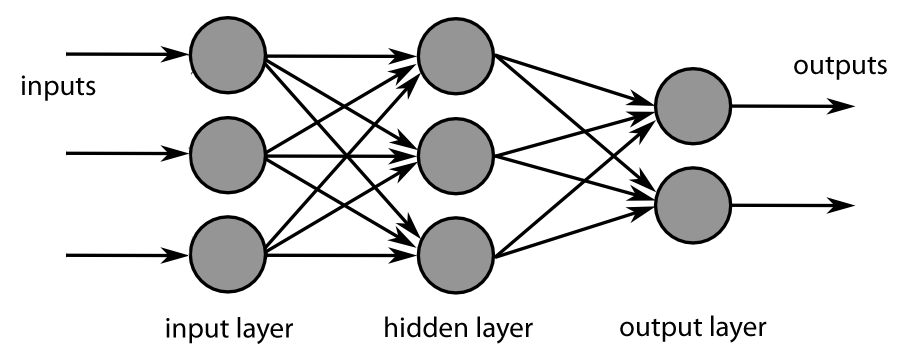}	        \captionsetup{justification=centering}
	\caption{Deep Neural Network}
	 \label{figure:dnn}
\end{subfigure}
\end{figure}

 Sentiment analysis from speech is a challenging topic which has been widely studied for years, but is still not applicable to real-time use cases in regard to speed and accuracy. First, we are going to further explore the problem with 3 different techniques:

1. Vocal feature extraction: Inspired by Dasgupta and Kwon et al.\cite{detection_human_emotion_2017_dasgupta,  Kwon2003}, we will extract vocal features (pitch, MFCC, formants, etc.) from raw speech signals and then feed them directly into classifiers such as SVM and HMM. \ref{figure:fe} shows the basic idea of vocal feature extraction.

2. Classification with segment-level features: Han et al. and Schuller et al. have shown that segment-level features are significant in speech sentiment analysis\cite{speech-emotion-recognition-using-deep-neural-network-and-extreme-learning-machine, acoustic_features_hybrid_schuller}. To further explore this idea, we will feed the raw audio segments and the processed audio segments (MFCC, STFT, etc.) into simple machine learning models such as SVM and ELM. \ref{figure:sfe} illustrates this technique.

3. Deep learning with context level analysis: With the rise of deep learning techniques, models that apply deep neural networks see a huge boost in accuracy \cite{speech_emotion_dcnn_pyramid_matching, adieu_features_end_to_end_dcrnn, emotion_cross_modal, neural_networks_powerful_function}. We will train our model on traditional deep neural networks (CNN, LSTM, etc.) as well as adapting some recent deep network variants such as DropConnect\cite{dropconnect} and ResNet\cite{resnet_kaiming_he} to see whether they can improve the accuracy in vocal sentiment analysis. \ref{figure:dnn} shows a typical deep neural network.

We will benchmark and improve these three methods and then try to combine them together for higher accuracy. We will also see if this problem can be solved in a real-time manner, as most use cases require the assistant to instantly resolve an user's mood and provide an appropriate response.

\section{Related works}
Sentiment analysis has conventionally been used in the realm of text data and corpuses. We will explore whether the sequential manner of speech data will also extend to prove to be helpful in classifying emotion.

The most common ways to detect human vocal emotion along with audio voice recognition are to use audio processing techniques such as MFCC, along with vocal filters like STFT (short time fourier transform). These features have to be handcrafted; and statistical analysis is used to form conclusions. According to Moore's law, computing speed became faster and faster due to the exponential number of transistors that can fit on a board grew, so did computing speed. Because of the massive parallelization and the nature of matrix operations, GPU/GPGPU and Google TPU evolved to became more widely adopted as the de facto way to use ``deep'' approaches. 

\paragraph{Defined Features}
\citet{detection_human_emotion_2017_dasgupta} has shown that algorithmic approaches for vocal sentiment analysis using pitch, timbre, SPL, and gap between words, rudimentary features can be extracted from vocal data and a range comparison could be done. Whether the differences are significant could be debated. 
\citet{perez-rosas_utterance-level_2013} has shown that features like prosody, voice, MFCC, spectral, etc, proves promising in identifying sentiment.
Conventional audio recognition approaches like MFCC and STFT (short time fourier transform) worked as a defined approach to deal with categorizing data. MFCC applies a FFT followed by a $y = \log(mag(x))$ and that output is applied thorugh a mel filter followed by another FFT, where the mel filter is given by $mel(f) = 2595 \log(1 + \frac{f}{700})$ where $f$ is given in hz. This allows for coefficients to represent ``bins'' which these coefficients belong to, the bands or bins represent the likely frequency domain values which are closer to human auditory range.
\citet{Kwon2003} has shown that using SVM and HMM (hidden Markov
model) to process pitch, log energy,
formant, mel-band energies, MFCCs and velocity/acceleration of pitch can receive a 96.3\% accuracy for 2-class classification, and 70.1\% for 4-class classification. They also proved that pitch and energy contribute most among all features. However, the result for 5-class classification is much lower.

For all the referenced algorithms, the time efficiency is good enough for small systems, but when it comes to classification for 4 categories or more, all approaches see a huge drop in accuracy and thus the usability for real-world user cases is arguable.

\paragraph{Context-free Approaches}

Data driven approaches became more prevalent after two milestones in technology: the explosion of data and rising power of computation. With the rise of social media and massive websites, data is more readily available and is able to be utilized by massive computing clusters. The implications of this gave rise to ``deep'' approaches to handling this data, particularly in form of neural networks. \citet{resnet_kaiming_he} proved with  Residual Neural networks that as layers grow linearly in a conventional CNN (convolutional neural network), vanishing gradients quickly become a problem. However this was rectified with a $$y = \mathcal{F}(x) + x$$, where $x$ and $y$ is defined as a layer's input and output and $\mathcal{F}(x)$ is defined as a forward pass on the previous layer. By adding the previous layer to a node's forward pass, essentially a layer's ``weight'' is increased and vanishing gradients is remedied through skipping. Emotion level classification was not explicitly tested in this paper, but results seem promising based on current performance and the same approaches could be used for training deep neural networks on speech recognition tasks. \\
\citet{speech-emotion-recognition-using-deep-neural-network-and-extreme-learning-machine} took an approach that recognized emotion by producing an treating output as an emotion state probability distribution. It extracted segments using MFCC to extract segment level utterances, and trained these utterances in a DNN network. This approach seems promising, however, using MFCC as an utterance level extractor may not always prove to be accurate. Additionally, this approach may  introduce bias through emotions with words that have a drawn out speech style, which skews the results of the MFCC through STFT. However, neural networks are powerful function approximators, and work well with data that does not have apparent features attached to it \cite{neural_networks_powerful_function}.

\citet{emotion_cross_modal} has used a pretrained video emotion classifier in order to reduce dependencies in labels while training the audio dataset. This approach uses the context learned from one neural network trained in video emotion recognition $f$ and uses that context to apply in another domain (audio) $g$ using cross-modal distillation. Additionally, this approach does not require labeled auditory data, which allows one network to teach the other; this model scales due to the fact one can essentially generate labels across domains. Training an additional neural network definitely incurs training time, but improves in parallelization through distillation.

\section{Solutions}

\subsection{Datasets}
For the majority of our testing and training, we were using the RAVDESS dataset \cite{RAVDESS_dataset} which consists of 8 classes of human emotion. This dataset contains labeled files in 3 modality (full AV, video-only and audio-only) and 2 vocal channels (speech and song) from male and female actors. Since our focus was on speech sentiment analysis, and also due to file size constraints, our models were trained on audio-only speech samples which consist of \begin{center} Labels: total=1440, neutral=96, calm=192, happy=192, sad=192, angry=192, fearful=192, disgust=192, surprised=192. \end{center} These labels correspond in order to the confusion matrix in \ref{figure:confuse}. Additional training and testing were done on the TESS dataset \cite{TESS} which contains a set of 200 sentences spoken by two actresses in 7 emotions. The dataset contains \begin{center} Labels: total=2800, neutral=400, happy=400, sad=400, angry=400, fearful=400, disgust=400, surprised=400. \end{center} Compared to the previous dataset, the TESS dataset is cleaner in loudness and length which brings more consistency to training and testing. However, all the sentences in this dataset are led by the phrase "Say the word", which can be less universal and lack variety, which may eventually cause overfitting and thus harm the final performance. Hence TESS here was only applied as an additional dataset.

All the data were  preprocessed and standardized sample by sample, as there were not too many samples, and standardizing across samples would make the data have little variance and hard to distinguish across classes. The normalization was based on loudness and length, for which we did zero padding to the heads and tails of all samples to make them length invariant.

\subsection{Architecture}
When testing the models, the training and data-processing were along the guidelines of   \citet{lee-tashev-high-level-feature-representation-using-recurrent-neural-network-for-speech-emotion-recognition}. For their features, they used 13-50 MFCC's along with 3 other features like pitch, voice probability, zero-crossing rate, and their first order derivatives. Derivatives and non-MFCC features were found to be mostly superfluous and that MFCCs captures the important features of the spectrogram. Our further tests found that although adding a large number of MFCC may cause some models to perform better, 13 MFCC's were sufficient in getting meaningful features for the model. Specifically during train processes, the model used either the flattened 13x26 MFCC window or the direct 2D window.

\begin{figure}[H]
\caption{}
\vspace{-5cm}
   \begin{subfigure}[b]{0.5\textwidth}\vspace{5cm}
  \centering
  \includegraphics[width=\textwidth]{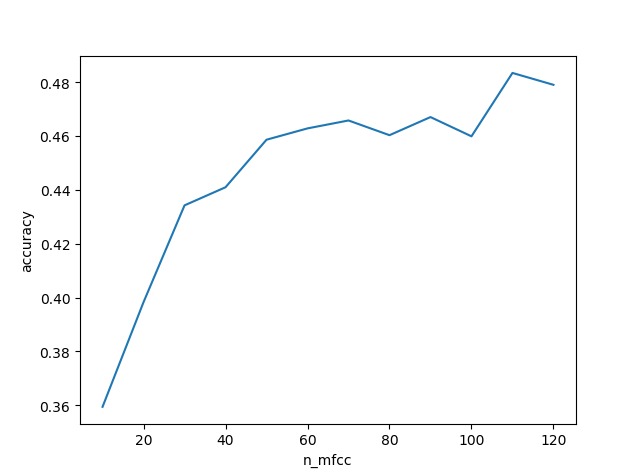}
  \caption{Result Using RBF}
    \label{figure:svm_mfcc_rbf}
\end{subfigure}
\begin{subfigure}[b]{0.5\textwidth}\vspace{5cm}
    \centering
\includegraphics[width=\textwidth]{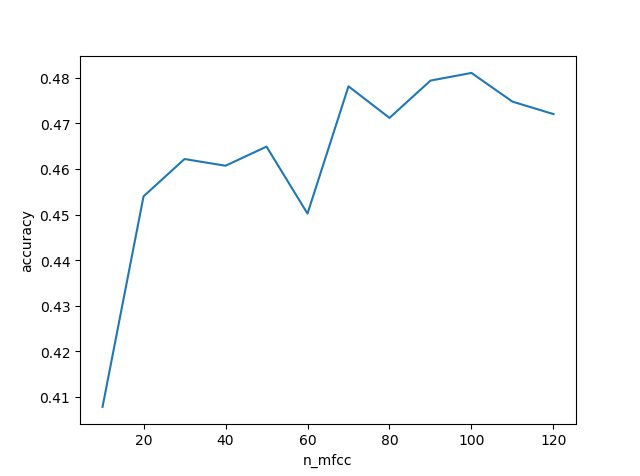}
  \caption{Result Using Linear}
    \label{figure:svm_mfcc_linear}
\end{subfigure}
\label{figure:svm}
\end{figure}

The first model we used was SVM, which performed reasonably well as a baseline test. We benchmarked the model with 2 different SVM kernels, RBF and Linear, with $C=10$ and $gamma=Scale$. And we also performed a parameter sweep on the number of MFCC, which ranged from 10 to 120, to test how significant the number of MFCC would affect the accuracy. We ran 10 times for each parameter and took the average, which yielded the result in \ref{figure:svm}. For both RBF and Linear models, the accuracy has a positive correlation with the number of MFCC, while it increases slower when the number of MFCC exceeds 20. However, when more than 50 MFCCs were applied, the accuracy started to fluctuate. Thus the current svm model parameters were found to be significant. For comparison concern, the best accuracy we could get for SVM is 48.11\%, which is achieved by using Linear kernel and 100 MFCCs.

\begin{table}[]
\centering
\caption{Model architecture used for CNN Model}
\label{table:cnn}
\begin{tabular}{@{}lll@{}}
\toprule
Layer (type)                   & Output Shape       & Param \# \\ \midrule
conv2d (Conv2D)                & (None, 13, 26, 32) & 320      \\
conv2d\_1 (Conv2D)             & (None, 11, 24, 32) & 9248     \\
max\_pooling2d (MaxPooling2D)  & (None, 5, 12, 32)  & 0        \\ \midrule
dropout (Dropout)              & (None, 5, 12, 32)  & 0        \\
conv2d\_2 (Conv2D)             & (None, 5, 12, 64)  & 18496    \\
conv2d\_3 (Conv2D)             & (None, 3, 10, 64)  & 36928    \\ \midrule
max\_pooling2d\_1 (MaxPooling2D)  & (None, 1, 5, 64)   & 0        \\
dropout\_1 (Dropout)           & (None, 1, 5, 64)   & 0        \\
flatten (Flatten)              & (None, 320)        & 0        \\ \midrule
dense (Dense)                  & (None, 512)        & 164352   \\
dropout\_2 (Dropout)           & (None, 512)        & 0        \\
dense\_1 (Dense)               & (None, 8)          & 4104     \\ \midrule
Total params: 233,448          &                    &          \\
Trainable params: 233,448      &                    &          \\
Non-trainable params: 0        &                    &          \\ \bottomrule
\end{tabular}
\end{table}


For the next model, convolutional neural networks were used to capture spatial information about the features. A standard CNN architecture along the lines of Alexnet \cite{alexnet} was used, commonly used for image recognition. Please see Table \ref{table:cnn} for more information about our specific layers and see Figure \ref{figure:nn} for a visual reference. 13x26 MFCCs were used for all of our tests and standardized each window by sample. Because CNN captures spatial data and MFCC has coefficient information along the x-axis, and its value on the y-axis, capturing correlation with convolution will provide a reasonable approach. Tests revealed that fitting a deeper neural network on our dataset often resulted in overfitting, especially with BatchNormalization layers \cite{batch_norm_ioffe} or Residual Blocks\cite{resnet_kaiming_he}, so this model architecture was suitable for our usages. Using approaches from literature, two convolutional blocks seemed suitable for the tasks as the features in MFCCs features are not particularly deep or ingrained. These layers were proceeded by two fully connected layers and one softmax layer with categorical crossentropy loss and accuracy as our training metric. The optimizer chosen was RMSProp with paremeters $lr = 0.0001$ and $decay = 1\mathrm{e}{-6}$ because as this is mostly categorical data, it makes sense to use RMSProp. 

\paragraph{Results}We reported a top accuracy of $85\%$ accuracy Top-1 (Top N refers predicted classes among the top N prediction choices) accuracy over 500 epochs using a 60/20/20 split on the testing data, showing that our approach was better than the 3-layer DNN architecture used by \citet{lee-tashev-high-level-feature-representation-using-recurrent-neural-network-for-speech-emotion-recognition}. We suspect this is because 2D convolution captures both the time domain from the MFCC and each frequency band from the 2D convolution. In addition to this approach, 0 padded 1D Convolution was also trained on the same task, giving similar accuracy, along with a Top-1 averaged anger score of nearly 90 percent.  The task of recognizing features from a MFCC is not unlike image recognition, so the results were not unexpected. CNN therefore provides an excellent way to recognize time series and frequency specific information due to the way that MFCC's store frequencies.

\section{Experiments}

\subsection{Model Adjustments}
To further improve accuracy, the initial model was tweaked slightly. A hyperparameter sweep and different layers from literature were added, including Residual blocks and skip connections and BatchNorm layers \cite{batch_norm_ioffe, resnet_kaiming_he}. These improvements did not improve the top level performance, due to the fact that the data is mostly low variance, and we did not have a problem of distinguishing features. A deeper neural network was also proposed during this process; the problem of fitting the low variance features was once again encountered. MFCCs give a limited amount of depth information for the data, and more features may needed to be added, such as pitch, prosody, and other segment level features,  \cite{lee-tashev-high-level-feature-representation-using-recurrent-neural-network-for-speech-emotion-recognition, perez-rosas_utterance-level_2013, speech-emotion-recognition-using-deep-neural-network-and-extreme-learning-machine}.

\subsection{Dataset Augmentation}
To rule out the possibility of overfitting, the dataset was augmented and the model was retrained with different parameters to increase sample size and sample variance. Firstly, the model was trained on the RAVDESS dataset with top level accuracy of 65 percent using the CNN model. After resolving overfitting issues a model was trained with reversed audio forms and their invertions (polarity flip). The combined dataset performed similarly as before, showing that data augmentation does not help as much as it does for image recognition tasks. This could be due to the fact that MFCC's discard some of the spatial information from the dataset, and collects simply the averaged 13 MFCC's from each feature window.
\section{Discussion}

After obtaining results, metrics and tuning was performed in order to ensure the best model architecture for our use case. 
The results reported below consist of the best tuned model. The model was obtained by hyperparameter tuning and augmentation as mentioned previously. 
\subsection{Metrics and results}

\begin{figure}[H]
\caption{}
\vspace{-5cm}
   \begin{subfigure}[b]{0.5\textwidth}\vspace{5cm}
  \centering
  \includegraphics[width=\textwidth]{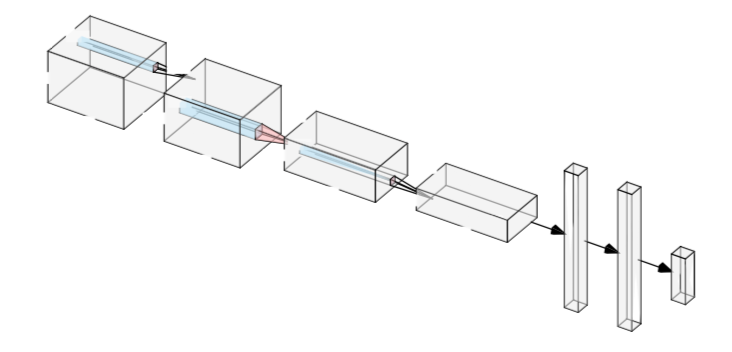}
  \caption{Neural network architecture}
    \label{figure:nn}
\end{subfigure}
\begin{subfigure}[b]{0.5\textwidth}\vspace{5cm}
    \centering
	\scalebox{0.5}{\input{confusion_matrix_cnn.pgf}}
	\captionsetup{justification=centering}
	\caption{Confusion matrix from running our averaged CNN on the RAVDESS dataset}
	 \label{figure:confuse}
\end{subfigure}
\end{figure}

\paragraph{Model evaluation} For metrics, precision, recall, and f1-score, and roc along with the confusion matrix (see figure \ref{figure:confuse}) were used to determine which classes were accurately tested and which weren't. The lower accuracy over the test on average is due to the possible data augmentation we performed in order to make them length invarient, as this adds no noise to the data, but still adds classification information. This is due to the fact that calm and neutral emotions both have relatively monotone samples. The top 3 most accurate classes were angry  ($86.8\%$), disgust($78\%$) , calm($72\%$). Intuitively, these classes produce wave forms which ``stand'' out more due to their waveform. The neutral class and calm class are relatively similar, so it is possible the model was not able to distinguish between the two. The least accurate classes are calm ($55\%$ accuracy 0.53 recall), neutral ($64\%$ accuracy, 0.79 recall) and happy ($66\%$ precision, 0.78 recall). It seems with these classes that although it was not able to distinguish the class from each other given the relatively lower precision and higher recall values. The results imply there is an overfitting issue or the dataset is not a good representation of the emotion state. This is similar to the results established by \citet{speech-emotion-recognition-using-deep-neural-network-and-extreme-learning-machine} where MFCCs cannot capture some long drawn utterances.

The biggest challenge over this project was feature extraction and choosing how to augment the data, and whether or not to standardize across samples or for their own sample. Sound features for vocal extraction are often used in the raw form for many voice recognition tasks like language translation, but for the uses of this project, MFCC's were helpful. Data augmentation for this particular task would not prove to be helpful in generating more data because of the fact reversed sound does not make spatial sense as it does with video and images. Training the model was also difficult at first, failing to get baseline accuracies produced by the SVM, but the hyperparmeter sweeps and testing a variety of models provided the framework for better model selection.

\begin{figure}[H]
\caption{}
\vspace{-5cm}
    \begin{subfigure}[b]{\textwidth}\vspace{5cm}
    \centering
    \includegraphics[width=\textwidth]{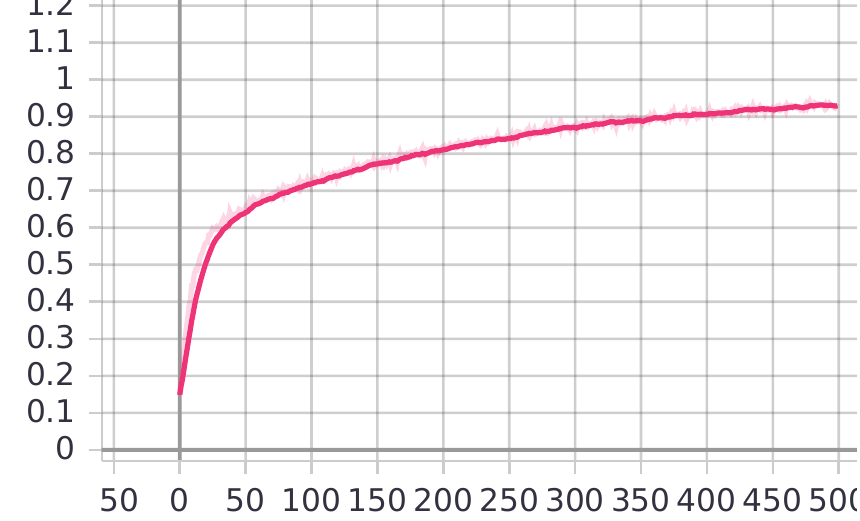}
	\caption{Epoch training accuracy (RAVDESS+TESS)}
    \label{figure:epoch_acc}
\end{subfigure}
\begin{subfigure}[b]{\textwidth}
    \centering
    \includegraphics[width=\textwidth]{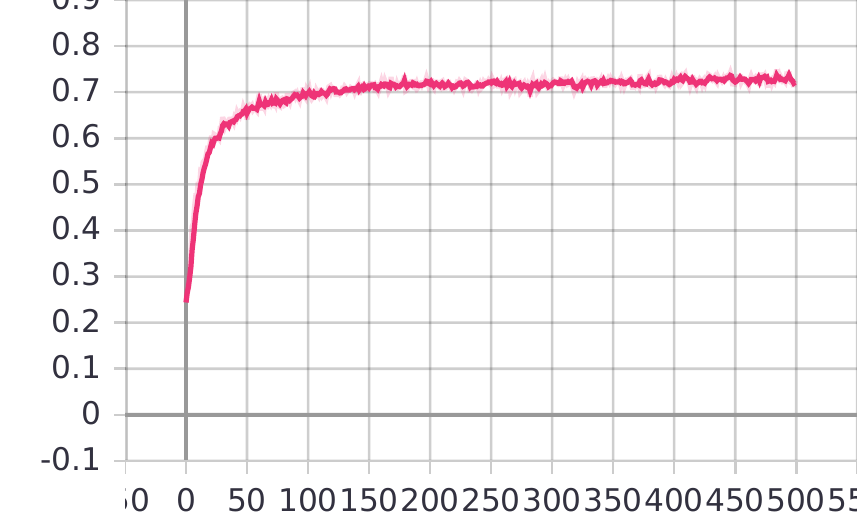}
	\caption{Epoch validation accuracy (RAVDESS+TESS)}
	\label{figure:epoch_val_acc}
\end{subfigure}
\end{figure}

\section{Conclusion}

\subsection{Future work}

For future work, we may choose to extend more features into our framework, but we did not find it necessary to add residiual layers to our networks when our task does not have too many labels, and the chance for overfitting is high. We can also choose to use to use bidirectional LSTM for training over audio datasets / MFCC. Additionally we may want to try an embedding framework, as attention frameworks seem to work well for many voice recognition tasks. Additionally with more training data and computational power, we would ideally have an ensemble of models so that we can get the best statistical properties over the dataset. Optimally, we also collect more data in the feature as TESS and RAVDESS only provide limited samples of user information \cite{TESS, RAVDESS_dataset}. We could have also tried more data augmentation such as stretch but most of those features would not be impacted for the FFT besides a phase shift. Overall, this project had a good direction to provide a framework for accurate human voice emotion recognition. For the problem of real time translations, models like Mobilenet will need to be explored for it's relatively small number of parameters and small architecture for real-time inference, as our relatively heaviweight CNN with large number of model parameters cannot provide an accurate and fast real-time response to spoken messages.

\subsection{Summary}
To summarize, MFCC's on audio data provide rich feature content about their data in a fixed manner, compared to raw audio wave forms. SVM's provide a fast way to fit data with non-linearities such as MFCC data. But because of the limitations of how they separate non-linear data, Convolutional neural networks capture rich features about their dataset with a high accuracy with the use of a large amount of data, and the same methods applied to photo recognition can also be applied to MFCC filters. With a smaller and more condensed model, a framework for real-time emotion recognition could be achieved. Machine learning for language recognition tasks seems very promising in the future, as computational speed and more larger sources of data become widespread.

\section{Acknowledgements}

We would like to thank Professor Keith Ross for supervising us throughout this project, Professor Xianbin Gu for mentorship and assistance, and Professor Olivier Marin for leading the capstone course. Our code will be available for public use and improvement.


\bibliographystyle{IEEEtranN}

\end{document}